Friday, July 01, 2016

# Nonlinear Dynamic Force Spectroscopy

Oscar Björnham[1] and Magnus Andersson[2,3]

[1]Swedish Defence Research Agency (FOI), SE-906 21 Umeå, Sweden, [2]Department of Physics, [3]Umeå Center for Molecular Research, Umeå University, SE-901 87 Umeå, Sweden

## Abstract

Dynamic force spectroscopy (DFS) is an experimental technique that is commonly used to assess information of the strength, energy landscape, and lifetime of noncovalent bio-molecular interactions. DFS traditionally requires an applied force that increases linearly with time so that the bio-complex under investigation is exposed to a constant loading rate. However, tethers or polymers can modulate the applied force in a nonlinear regime. For example, bacterial adhesion pili and polymers with worm-like chain properties are examples of structures that show nonlinear force responses. In these situations, the theory for traditional DFS cannot be readily applied. In this work we expand the theory for DFS to also include nonlinear external forces while still maintaining compatibility with the linear DFS theory. To validate the theory we modeled a bio-complex expressed on a stiff, an elastic and a worm-like chain polymer, using Monte Carlo methods, and assessed the corresponding rupture force spectra. It was found that the nonlinear DFS (NLDFS) theory correctly predicted the numerical results. We also present a protocol suggesting an experimental approach and analysis method of the data to estimate the bond length and the thermal off-rate.

# Introduction

In the late nineties, Evans et al. formulated the theoretical basis for dynamic force spectroscopy (DFS) (Evans and Ritchie 1997; Merkel et al. 1999) that could allow for a time varying force in the expression for bond dissociation originally given by Kramers (Kramers 1940), and later refined by Bell (Bell 1978). Since then, DFS has successfully been employed in a wide range of applications in the field of biophysics to assess information of the strength, energy landscape, and lifetime of bio-molecular interactions (receptor-ligand bonds) at the molecular scale. DFS is commonly performed using force transducers such as Optical Tweezers (OT) or Atomic Force Microscopy (AFM) instruments, which can apply loading rates from a few pN/s to several nN/s. Using these techniques and the theory for DFS, atomic information of, for example, the Streptavidin-Biotin complex (Lee et al. 1994; Yuan et al. 2000), the digoxigenin-antibody complex (Neuert et al. 2006), the Mucin 1 – antibody bond (Sulchek et al. 2005), and the unbinding force of complementary DNA (Strunz et al. 1999) have been revealed.

Although the theory is very useful in its current state, a restriction of the DFS theory is the requirement of a force that increases linearly with time. An experimental system can often be configured to provide an external force that meets this constraint for a limited force interval. However, for some complexes under study there can be situations when a more general time dependency of the external force in the DFS theory is required. For example, when the rupture force surpasses the linear span of the probe in a force spectroscopy apparatus or when measurements on bonds connected to biological tissues or organelles that have an intrinsic nonlinear response to external forces are investigated.

Many bacterial adhesion organelles, commonly called pili, expressed by uropathogenic, enterotoxigenic, and respiratory tract associated bacteria exhibit nonlinear force responses (Forero et al. 2006; Miller et al. 2006; Andersson et al. 2007; Chen et al. 2011; Castelain et al. 2011; Mortezaei et al. 2015a; Mortezaei et al. 2015b). In particular extensions of pili expressed by uropathogenic and enterotoxigenic bacteria, show distinct nonlinear force responses, i.e., they respond to an external force initially by unwinding the structure at a constant force and thereafter by a pseudo-elastic force response. A nonlinear force response, similar to that of a worm-like chain (WLC) polymer, is also seen when extending T4 adhesion pili expressed by *Streptococcus Pneumonia* (Castelain et al. 2009). The response from these adhesion organelles are thus nonlinear with extension by nature. To assess information of the adhesin using DFS, the theory needs to be refined.

In this work, we extend the DFS theory to include nonlinear external forces. The theory, referred to as nonlinear dynamic force spectroscopy (NLDFS), covers positive loading rates in a nonlinear regime where the force increases continuously. NLDFS is compatible with the linear DFS theory, which is shown to be a special case. We validated NLDFS by modeling an adhesin expressed on polymers that exhibit different nonlinear force responses using Monte Carlo simulations, and by assessing the rupture force spectra for various extension velocities. The results, for all tested cases, show that the nonlinear DFS theory correctly predict the most probable rupture force obtained from the stochastic modeled data and that the methodology suggested in this work is applicable to a variety of experimental investigations.

# Theory

## Linear dynamic force spectroscopy

We start by briefly introduce the traditional DFS theory. For detailed information see (Evans 2001). In 1940 Kramers (Kramers 1940) used the Smoluchowski equation to conclude that the transition rate, i.e., the thermal off-rate $k_{off}^{th}$, which gives information of how frequently a bond transits between two states separated by an energy barrier $E_T$, can be described by an Arrhenius factor,

$$k_{off}^{th} = \nu_a e^{-\frac{E_T}{k_B T}}, \tag{1}$$

where $\nu_a$ is the attempt rate originating from the molecular vibrations in an overdamped condensed system, $k_B$ is the Boltzmann´s constant, and $T$ is the absolute temperature. Almost 40 years later Bell (Bell 1978) inserted the effect of an external force, $F$, to the theory of Kramers to obtain an expression for the force-dependent dissociation rate $k_{off}$, given by

$$k_{off} = k_{off}^{th} e^{\frac{F x_b}{k_B T}}, \tag{2}$$

where $x_b$ is the bond length that describes the spatial distance between the energy minima and the transition energy barrier. Evans et al. extended Bells work by including a linearly increasing force described by a constant loading rate defined as the time derivative of the force, $r$ (1-3). They showed that with increasing force the probability for the bond to rupture increases. This implies that there will be a maximum likelihood for bond rupture for a specific force. This force, which often is referred to as the *most probable rupture force*, $F^*$, depends on the bond length, the thermal off-rate, and the loading rate and is found as the peak force in a rupture force spectrum. We will refer to this entity as the *peak force* throughout the rest of this work. Evans et al. showed that the peak force can be explicitly expressed as

$$F^* = \frac{k_B T}{x_b} \ln \left( \frac{r x_b}{k_{off}^{th} k_B T} \right). \tag{3}$$

## Nonlinear dynamic force spectroscopy

To reduce the complexity when analyzing experimental data obtained from nonlinear loading rates we restrict the range of the nonlinear theory to continuous forces with positive time dependent loading rates. This implies that the force is always increasing and that every force value is only present once. This restriction is not effectively limiting the usability of NLDFS since in a force spectroscopy experiment the system under study is not likely to be exposed to external forces with alternating negative and positive loading rates.

To derive the necessary equations for NLDFS we start by consider the probability $P$ of an intact bond. The probability rate of bond rupture equals the negative change of $P$ over time and can be expressed as

$$-\frac{dP}{dt} = -\frac{dP}{dF} r = k_{off} P \, , \tag{4}$$

where $r$ is the loading rate defined as

$$r \equiv \frac{dF}{dt} \, . \tag{5}$$

The derivative of the probability rate, which is indicative of the position of the peak force, can thereby be expressed in two ways, wiz. as

$$\frac{d^2 P}{dt^2} = -\frac{d(k_{off} P)}{dt} = -\frac{dk_{off}}{dt} P - k_{off} \frac{dP}{dt} = P \left( k_{off}^2 - \frac{dk_{off}}{dt} \right) \tag{6}$$

and as

$$\frac{d^2 P}{dt^2} = \frac{d}{dt} \left( \frac{dP}{dF} r \right) = \frac{d}{dt} \left( \frac{dP}{dF} \right) r + \frac{dP}{dF} \frac{dr}{dt} = \frac{d}{dF} \left( \frac{dP}{dF} \right) r^2 + \frac{dP}{dF} \frac{dr}{dt} \, . \tag{7}$$

The term

$$\frac{d}{dF} \left( \frac{dP}{dF} \right) r^2 \, ,$$

is zero at the peak force, which implies that it is possible to combine Eqs. (6) and (7) to obtain the following relationship at the peak of the force rupture spectrum,

$$\frac{dP}{dF} \frac{dr}{dt} = P \left( k_{off}^2 - \frac{dk_{off}}{dt} \right) \, . \tag{8}$$

Thus by inserting the expression for $dP/dF$ from Eq. (4) into Eq. (8), we obtain

$$\frac{1}{r} \frac{dr}{dt} = \frac{1}{k_{off}} \frac{dk_{off}}{dt} - k_{off} \, , \tag{9}$$

which, in turn, can be written as

$$\frac{d \ln r}{dt} = \frac{d \ln k_{off}}{dt} - k_{off} \, . \tag{10}$$

Making use of the expression for the dissociation rate, Eq. (2), gives the resulting relation between the loading rate and the peak force as

$$\frac{d \ln r}{dt} = \frac{r x_b}{k_B T} - k_{off}^{th} e^{\frac{F^* x_b}{k_B T}} \, , \tag{11}$$

which can be reformulated as

$$F^* = \frac{k_B T}{x_b} \ln\left[ \frac{1}{k_{off}^{th}} \left( \frac{r x_b}{k_B T} - \frac{d\ln r}{dt} \right) \right]. \tag{12}$$

## Wormlike chain model

We used the WLC model as a case study to evaluate the NLDFS theory. The WLC model is commonly used to describe the nonlinear entropic driven force response of biopolymers exposed to external forces (Strick et al. 2002; Kiss et al. 2006; Bianco et al. 2007; Björnham et al. 2008; Björnham and Schedin 2009). In this model the force can be expressed as a function of the distance between the two ends of the polymer, which, for the inelastic case, is given by

$$F = \frac{k_B T}{l_p} \left[ \frac{1}{4}\left(1 - \frac{L}{L_c}\right)^{-2} - \frac{1}{4} + \frac{L}{L_c} \right], \tag{13}$$

where $l_p$ is the persistence length, $L$ is the Euclidian distance between the two ends of the molecule, and $L_c$ is the contour length of the polymer. The contour length is the structural length of the polymer and equals $L$ if the polymer is fully stretched. If the polymer is extended at a constant velocity, $v$, the parameter $L$ can be expressed as

$$L = vt. \tag{14}$$

This implies that the effective loading rate can be given as

$$r(t) \equiv \frac{dF(t)}{dt} = \frac{dF}{dL}\frac{dL}{dt} = \frac{k_B T}{l_p} \frac{v}{L_c} \left[ \frac{1}{2}\left(1 - \frac{vt}{L_c}\right)^{-3} + 1 \right]. \tag{15}$$

The NLDFS theory then predicts that the peak force should be given by Eq. (12) with $r(t)$ being given by Eq. (15) for WLC.

# Results and Discussion

## Validation of the NLDFS theory

To investigate the validity of the NLDFS theory for different time-dependent external forces, it was compared to the analytic solution for the rupture probability, i.e., Eq. (4) together with Eq. (2). Further on, numerical simulations by means of Monte Carlo (MC) methods where conducted and used as validation. We set the bond length, $x_b$, to 0.70 nm and the thermal off-rate, $k_{off}^{th}$, to $10^{-4}$ s$^{-1}$, which are values in the typical range for noncovalent adhesion bonds (Sulchek et al. 2005; Björnham et al. 2009). The bio-complex was exposed to three different force responses all with an extension velocity of 10.0 µm/s. To verify the simulation and the analysis procedure, three sets of data with one million measurements each, were compiled using a narrow Gaussian kernel with a width of 0.50 pN, see Figure 1. The Gaussian kernel function will push the peak of the distribution slightly towards lower forces since the analytic distribution is skewed, which in turn, will result in a net flow of probability density towards lower forces at the peak as the kernel is applied. Although the effect is negligible here, it is recommended, in both DFS and NLDFS, to carefully choose the width of the Gaussian kernel to minimize this effect and at the same time obtain a smooth curve to identify the peak force.

In the first case, the force was increased linearly with time, thus resulting in a constant loading rate. This implies that Eq. (12) is reduced to Eq. (3), which is the commonly used expression for the peak force in linear DFS. For the case with a loading rate of 100 pN/s, Fig 1A shows the resulting rupture force probability spectra. The inset shows, qualitatively, the time evolution of the force. As can be seen, the analytical solution (black dashed line), which according to Eq. (12) is 70.7 pN, coincides perfectly with that of the Monte Carlo simulations (red line). Moreover, the predicted peak force (green vertical dashed line), using Eq. (12), matches the force for which the distribution has a maximum of both curves.

To model nonlinear increasing forces, e.g., to mimic cases when a receptor-ligand pair is attached to a membrane or polymer, we applied both a quadraticly increasing force (elastic reversible polymer) and a force that follows that of a WLC model, i.e., Eq. (13). Figure 1B and 1C, respectively, display the rupture probability densities from the simulations using these two nonlinear forces. The two panels show that the peak forces predicted by the theory agree with that of the simulations for both the case with a quadraticly increasing force, 52.3 pN, and for the case with the WLC, 41.6 pN. Since the peak forces predicted by the theory given above are in good agreement with those of the numerical solutions for all force curves, we conclude that the NLDFS theory can accurately predict the peak force of a receptor-ligand pair connected with polymers showing linear as well as nonlinear force responses.

Note that in Figure 1C, when bonds are linked via WLC polymers, the rupture probability curve shows two peaks. These two peaks can be explained by the initial slowly increasing force and the final rapidly increasing force experienced by bonds linked to WLC polymers that are extended. Thus, two effective loading rates are possible resulting in a small fraction of bonds breaking at the lower loading rate. The ones that persisted therefore break at the higher loading rate.

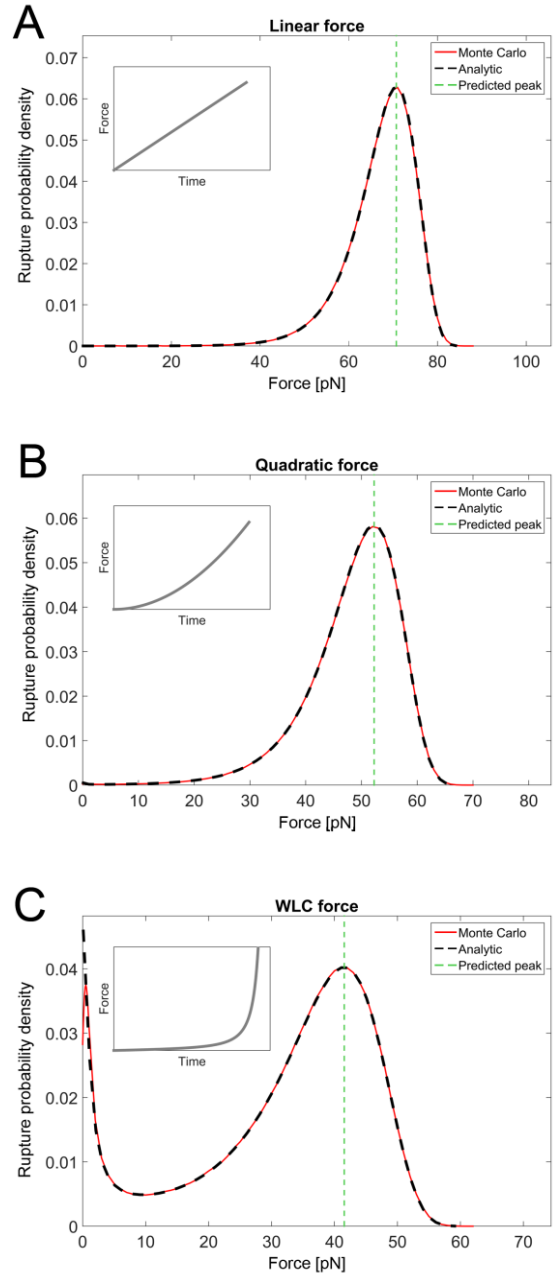

Figure 1. An example of the rupture probability distribution using one million samples for a velocity of 10.0 $\mu$ms$^{-1}$. The black dashed line is the analytical solution while the red line is the density estimate from the Monte Carlo simulations, using a Gaussian kernel with standard deviation of 0.50 pN. The vertical green line is the peak force, $F^*$, predicted by Eq. (12). The agreement is excellent except for a small deviation at the smallest forces in the WLC-case due to inherent properties of the kernel density estimation method at the boundary of the interval. The inset figure depicts the relation of the applied force with respect to time. For the linear and quadratic cases the force was given by $F = a_1 x$ and $F = a_2 x^2$ with the constants $a_1$ and $a_2$ set to 10 pN/$\mu$m and $10^{-3}$ pN/$\mu$m$^2$, respectively. For the WLC case the force was given by Eqs. (13) and (14) with $L_c = 10.0$ $\mu$m and $l_p = 3.00$ nm.

## Applying NLDFS theory to experimental data

In practice, however, it is of limited use to calculate the peak force solely for the cases when the bond length and the thermal off-rate are known *a priori*. Instead, the theory must be able to serve as a tool for experimentalists to estimate these two parameters from measurement data. The question is then: how do you design an experiment protocol to extract the desired parameter values?

### DFS

DFS provides a technique to assess the bond length and the thermal off-rate following a straightforward scheme. First, the bio-complex under study is exposed to an external force that increases linearly with time, i.e., with a constant loading rate. Eventually the bond will break, giving rise to a rupture force. However, a single rupture force is solely one sample from the probability distribution that represents the specific loading rate, the bond length, and the thermal off-rate. To obtain sufficient statistics to quantify the distribution, the rupture force must be sampled many times for a given loading rate. Second, the rupture force spectrum is constructed from the set of rupture forces obtained, whereby the peak force is identified by localizing the peak of the distribution. Third, the loading rate is changed and a new set of measurements are conducted resulting in a new value of $F*$. Thus, for every loading rate, a corresponding value of the peak force is given. The bond length and the thermal off-rate can thereafter be found by fitting Eq. (3) to this set of data. This is a well working method that has been widely used. However, as was alluded to above, a limitation of DFS is the constraint that Eq. (3) is valid only under the assumption of constant loading rates.

### NLDFS

As described above, DFS utilizes constant values of the loading rates in the experiments. These values of loading rates are used in pair with their corresponding values of the most probable rupture forces to find estimates of the bond length and the thermal off-rate.

In NLDFS, the loading rate is not constant during a measurement, which therefore requires a slightly different approach. Instead of keeping the loading rate constant, the pulling velocity, $v$, of the force transducer is held constant during a measurement. Hence, the rupture forces are recorded as for DFS but the peak force is paired with the corresponding velocity. To find the peak force value that corresponds to this pulling velocity, the velocity is then changed and a new set of measurements is performed. Thus, for each velocity there is a corresponding value of the peak force. These data are used together with Eq. (12) to obtain the estimated parameter values. However, to do this, an expression for the loading rate as a function of the velocity needs to be derived. For the case when the force only depends on the position, i.e., $F = F(L)$, this can be done in the following way. First, it should be noted that the loading rate can be written as

$$r \equiv \frac{dF}{dt} = \frac{dF}{dL}\frac{dL}{dt} = \frac{dF}{dL}v \qquad (16)$$

where $L$ now is a measure of the position of the force transducer, in general given by $vt$ where both $v$ and $t$ are known entities. The derivative $dF/dL$ is in general a function of $L$ that needs to be known or assessed, which can be found either through theoretic consideration of the system or by using measurement data. When this relation is established, Eq. (12) provides a full prediction of the expected value of the peak force given the velocity, the bond length and the thermal off-rate. Even though Eq. (12) might turn out to be an implicit function, the solution for the peak force can readily be found. This means that for every combination of the parameter values there will be one theoretical and one measured value of the peak force. Standard algorithms may then by utilized to find which parameter values that minimize the mean square error of these forces for all velocities.

## Protocol for NLDFS

A general description of how to obtain the bond length and the thermal off-rate using measurements was shown above. We will here give a more explicit protocol how this could be done in practice. The procedure is based on Eq. (12). The loading rate needs to be formulated as a function of the velocity whereafter a fitting algorithm can be applied. For this we suggest the following approach:

1) Measure $F^*$ for different velocities $v$. It is possible to use only two different velocities but highly recommended that at least four different velocities are used to obtain better accuracy.
2) Find a relation between $F^*$ and $v$. Note that $r$ can be expressed as function of $v$ in an experiment, i.e.,
   a. Relate the loading rate $r$ to the velocity $v$ using Eq. 16.
   b. Use Eq. (12) to define a, possible implicit, relation between $F^*$ and $v$.
   Using a and b, there is a relation between $v$ and $F^*$ that depends only on $x_b$ and $k_{off}^{th}$.
3) The parameter values $x_b$ and $k_{off}^{th}$ can now be assessed using a standard fitting procedure with the coupled values of $v$ and $F^*$.

### Numeric example using a WLC

As a well-controlled example we numerically simulated a force spectroscopy experiment of a receptor expressed on a tip of a polymer with WLC properties that was bound to an immobilized ligand. This simulation thus mimicked an experiment using AFM or OT instrumentation. The parameter values of the WLC model were set as; bond length, $x_b$, 0.70 nm, thermal off-rate, $k_{off}^{th}$, $1.00 \cdot 10^{-4}$ Hz, persistence length, $l_p$, 3.00 nm, contour length, $L_c$, 10.0 µm, and thermal energy, $k_b T$, 4.11 pNnm. Since the elastic stiffness of the force probe, i.e., the AFM cantilever or the bead in the optical trap, is significantly higher than the elastic properties of the modeled WLC polymer, we modeled these as infinitely stiff. We thereafter analyzed all data closely following the approach described above:

1) The peak force $F^*$ was identified for four different velocities $v$.
2) We calculated $t$, to be used in Eq. (15), for every $F^*$ with the corresponding velocity $v$ by using Eqs (13) and (14). This relation between time and the force can also be readily measured during the experiments. With the time corresponding to the peak force and the loading rate function given by Eq. (15) we had everything we needed to use Eq. (12) as a relation between $F^*$ and $v$. This means that $F^*$ was expressed as a function of the velocity.

3) The acquired pair values for $F^*$ and $v$ were now used. A standard algorithm that finds the parameter values of the bond length and the thermal off-rate that minimizes the mean square error of the theoretical and measured values of $F^*$ was utilized.

For each of the four different extension velocities: 10, 100, 1 000, and 10 000 µm/s; 50 MC force spectroscopy simulations were performed. The rupture forces were saved and four continuous rupture probability density distributions, using a Gaussian kernel density estimator ($\sigma = 3\,\mathrm{pN}$), were generated. The peak force was identified for each of the four distribution. Figure 2A shows the rupture force spectrum for the highest velocity. To estimate the bond length and thermal off-rate we thereafter numerically fitted Eq. (12) to the data using a Nelder-Mead simplex algorithm to find the parameter values that minimized the mean square error of the peak forces. The data from the simulation are shown with the fitted values in Figure 2B and Table 1.

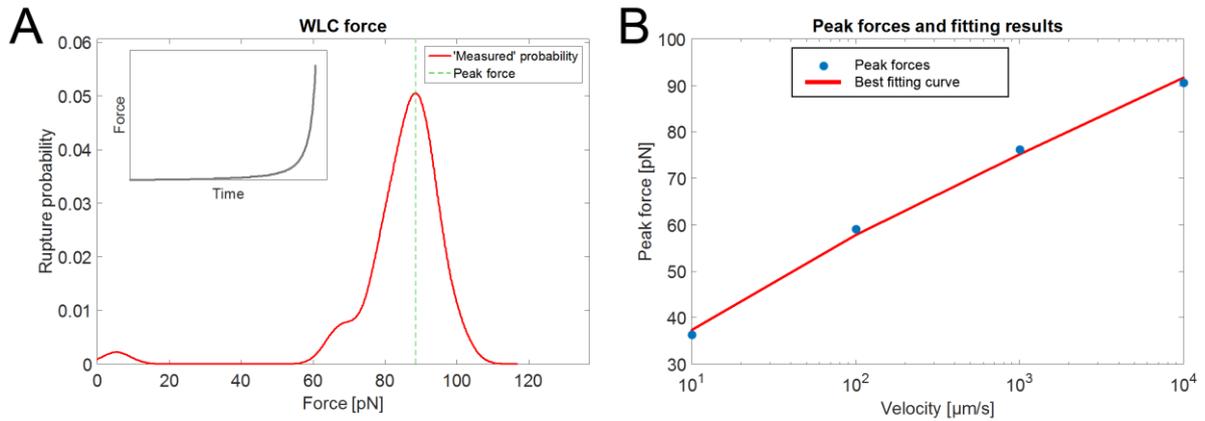

Figure 2. Panel A, a spectrum of rupture forces obtained for $v = 10\,000$ µms$^{-1}$ from 50 measurements. The peak force 88.47 pN. Panel B the best estimates of the bond length and thermal off-rate obtained from a fitting algorithm based on data from four velocities and the corresponding peak forces.

Table 1. The numerical results for the simulated example with the resulting values for the thermal off-rate and the bond length.

| $N$ | $v = 10\,$µm/s | $v = 100\,$µm/s | $v = 1\,000\,$µm/s | $v = 10\,000\,$µm/s | $x_b$ | $k_{off}^{th}$ |
|---|---|---|---|---|---|---|
| 50 | 39.57 pN | 55.33 pN | 74.58 pN | 88.47 pN | 0.685 nm | $1.54 \cdot 10^{-4}$Hz |

The assessed parameter values for this simulation are close estimates of the true values, where the bond length is underestimated with only ~2.1 %, whereas the thermal off-rate is overestimated by ~54 %. The discrepancy is expected due to the stochastic nature of the receptor-bond complex and should therefore depend on the sample size. Since we conducted the analysis based on only 50 simulations, we expect this error to decrease significantly with increased sample size. To quantify the error in the parameters we conducted a statistical analysis of the data (described below).

## Error dependencies of the sample size

In the example above, measurement sets consisting of 50 rupture forces were used to identify the peak forces for each velocity. Due to the stochastic nature of the experiment, the accuracy is expected to be improved with larger data sets, i.e., the more rupture forces that are sampled the less error there will be in the parameter values. To acquire acceptable accuracy of the parameter values it is in general recommended in the literature to conduct at least ~50-100 rupture measurements (Evans 1999; Merkel et al. 1999; Björnham and Schedin 2009). Therefore, we performed simulations with 50, 70, 100 and 300 rupture force, which allowed us to quantify the expected error in the parameters as a function of the sample sizes. In addition, a control set with ten million rupture forces for each velocity were performed. The resulting mean errors from these simulations for the bond length and the thermal off-rate are presented in Figure 3 and Table 2. Since experiments normally are conducted with 50-300 samples the mean relative error of the bond length is found to be less than 4 % while the mean relative error in the thermal off-rate is ~50%. This difference in errors can be explained by the fact that the rupture force is significantly more sensitive to the bond length in comparison to the thermal off-rate, therefore this difference is less remarkable.

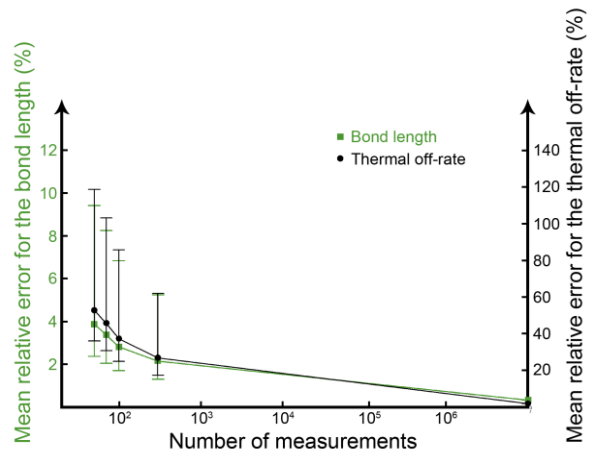

Figure 3. Mean relative error of the parameters as function of number of measurements. The error bars show the quartiles of the stochastic distribution of retrieved parameter values.

Table 2. Statistical measures of the relative errors in the resulting parameter values in comparison to the analytic values. The data was obtained by calculating the most probable rupture forces from $N$ measurements at four different velocities and the bond length and thermal off-rate were calculated using the method described in the theory section. This procedure were then repeated 10 000 times to quantify the expected errors in the parameter values.

| Method | Samples $N$ | Iterations | Mean relative error for the peak force [µm/s] | | | | Mean relative error | |
|---|---|---|---|---|---|---|---|---|
| | | | $v = 10^1$ | $v = 10^2$ | $v = 10^3$ | $v = 10^4$ | $x_b$ | $k_{off}^{th}$ |
| Monte Carlo | 50 | 10 000 | 5.42% | 3.13% | 2.26% | 1.81% | 3.88% | 52.7% |
| Monte Carlo | 70 | 10 000 | 4.79% | 2.77% | 1.99% | 1.59% | 3.37% | 45.5% |
| Monte Carlo | 100 | 10 000 | 4.17% | 2.38% | 1.76% | 1.41% | 2.81% | 37.3% |
| Monte Carlo | 300 | 10 000 | 3.06% | 1.73% | 1.28% | 1.01% | 2.15% | 26.7% |
| Monte Carlo | $10^7$ | 1 | 0.1% | 0.1% | 0.3% | 0.1% | 0.32% | 1.89% |
| | | | | | | | | |

| Analytical, most probable rupture forces [pN] | 41.55 | 58.61 | 74.49 | 89.81 | - | - |

## Conclusion

We have presented an extension of the standard DFS theory that can accommodate also nonlinear forces denoted NLDFS. The NLDFS theory enables investigation of a wide range of biomechanical systems that show nonlinear force responses without compromising with the well-established and frequently used linear DFS. Examples of receptor-ligand systems that can be analyzed using this theory are adhesins expressed on bacterial adhesion pili.

The data analysis using NLDFS requires a slightly more advanced fitting procedure than the conventional DFS theory to acquire the parameter values of the bond length and the thermal off-rate. The reason for this is that the loading rate becomes dynamic given by the introduction of a new term, $d \ln r / dt$ in Eq. (12). This extra term, however, disappears for constant loading rates which shows that the NLDFS theory reduces into the regular linear DFS for the case with a constant loading rate. In DFS, a set of different loading rates with the corresponding values of the peak forces are used. Finally, just as assumed in DFS experiment, we neglect the dynamic effects of the viscous drag force on the probe since the pulling velocities are slow.

To conduct the equivalent procedure in NLDFS the protocol has to be modified. Instead of keeping the loading rate constant during experiments, the pulling velocity is kept constant. If the force increases linearly with distance, the loading rate in Eq. (16) is constant and the NLDFS analysis falls into the linear DFS-regime. This implies that the velocity can be used as the entity kept constant in measurements using both DFS and NLDFS theory.

Evans et al. refined the concept of DFS by introducing soft polymers linking the receptor-ligand bond (Evans and Ritchie 1999). By defining a compliance function, they compared how the peak force changed with and without a soft linker. Their approach allows for analysis of bond strengths in the presence of nonlinear external forces by defining the polymer force response using a relation between the probe stiffness and the characteristic stiffness of the polymers; and utilizing an apparent loading rate, which equals a constant probe stiffness multiplied with the pulling velocity. A theoretical correlation can thereafter be established, which is used in a curve fitting procedure. The main concepts of that method and NLDFS presented in this work are similar. However, the NLDFS theory utilizes a more direct approach and introduces only a minimal modification of the linear DFS. Explicit information of the stiffness of the probe and the soft linker in the system can be readily bypassed by direct investigation of the force vs. distance curve, which also implies that nonlinear responses in the probe is treated equally as nonlinear responses in the bio-complex linker. In other words, in the approach presented here, only the force experienced by the bond under investigation is considered, disregarding the origin of the force response since it has no impact in the analysis.

Another approach to deal with nonlinear loading rates is proposed in reference (Friedsam et al. 2003). Instead of using the peak force values they used the probability density function for bond rupture. This expression is, however, rather complicated and depends on; the force, the loading rate, the bond length and the thermal off-rate, where the force and the loading rate distributions are found by investigating the experimental data. Estimates of the bond length and thermal off-rate may thereby be found by fitting the function to the rupture probability data. This method uses all data points, and not only the ones close to the peak force, which is of advantage since it make use of a larger data set. On the other hand, the method is sensitive to outliers and measurement artifacts.

Besides the theoretical framework presented here, a protocol for how to conduct a practical measurement and data evaluation is described by a numerical example using Monte Carlo simulations. Since the rupture forces are stochastic, the parameter values will inherently have uncertainties coupled to them. We quantified the expected uncertainties by a large number of iterative simulations that provided the magnitude of errors that one would expect in a real experiment. It was found that already at 300 experimental data points the mean relative error for the bond length is only ~2%. However, in a real experiment using AFM or OT, additional measurement error and noise will be added on top of the inherent stochastic nature of the bond under investigation. Thus, the values of the expected errors presented here are therefore to be interpreted as a best case outcome.

## Acknowledgements

This work was supported by the Swedish Research Council (2013-5379) and from the Kempe foundation to M.A.